\documentclass[aps,prl,superscriptaddress,twocolumn,
groupedaddress,nofootinbib,
nobalancelastpage,nobibnotes]{revtex4}

\usepackage{amsmath}
\usepackage{amssymb}
\usepackage{epsfig}
\usepackage{graphicx}

\newcommand{\bi}{\begin{itemize}}
\newcommand{\ei}{\end{itemize}}

\newcommand{\be}{\begin{equation}}
\newcommand{\ee}{\end{equation}}
\newcommand{\bea}{\begin{eqnarray}}
\newcommand{\eea}{\end{eqnarray}}

\newcommand{\oil}{\overline}

\begin{document}

\title{  Electron to Muon Conversion in
  Electron-Nucleus Scattering
  as a Probe of Supersymmetry
}

\author{T. Bla\v{z}ek
\footnote{{\footnotesize On leave of absence from
the Dept. of Theoretical Physics, Comenius Univ., Bratislava, Slovakia}}}
\email[E-mail address: ]{blazek@hep.phys.soton.ac.uk}
\affiliation{Department of Physics and Astonomy,
University of Southampton, Southampton, SO17 1BJ, United Kingdom}
\author{S. F. King}
\email[E-mail address: ]{sfk@hep.phys.soton.ac.uk}
\affiliation{Department of Physics and Astonomy,
University of Southampton, Southampton, SO17 1BJ, United Kingdom}


\begin{abstract}
\vspace*{0.2cm}
 We suggest that $e\to \mu$ conversion
 in low-energy electron-nucleus scattering is a new and potentially
 observable indirect signal of supersymmetry and should
 be searched for in experiment.
 We estimate the rate for this process in a straightforward
 calculation taking into account the existing constraint
 from non-observation of the $\mu\to e\gamma$
 decay and find the cross-section to be $\sigma < 10^{-8}\,$fb.
\end{abstract}

\pacs{???}

\vskip 0.1in
\noindent

\maketitle

\subsection{Introduction}
The recent advances in neutrino physics have established that not
only do neutrinos have mass, but also they mix strongly, thereby
violating the separate lepton numbers $L_e, L_\mu , L_\tau$
which, if preserved, would forbid such Lepton Flavour Violating (LFV)
processes as $\mu\to e\gamma$ \cite{lfv}.
On the other hand, in the simplest
extensions of the Standard Model, augmented to accomodate neutrino
mass and mixing, the rate of such LFV processes is exceedingly small,
being suppressed by the ratio of neutrino mass to W boson mass,
rendering $\mu\to e\gamma$ effectively unobservable.

In the minimal supersymmetric standard model (MSSM), extended to
accomodate neutrino mass and mixing, the situation
is dramatically improved since any LFV in the lepton sector
tends to become transferred to the slepton sector, via
radiative corrections, leading to unsuppressed LFV rates
which depend on ratios of superpartner masses. This has led to
a considerable amount of activity associated with LFV in
the MSSM and related models \cite{lfv}.

Much of the interest in LFV has focussed on limits arising from
the classical decay experiments such as $\mu\to e\gamma$.
Other experiments, usually referred to as $\mu - e$ conversion,
look for LFV arising from low energy muons being captured in
atoms, where the captured muon may subsequently convert to
an electron whose characteristic energy is slightly below the
muon mass \cite{Feinberg:1959ui}.

Recently an alternative approach to searching for LFV has been
proposed based on fixed-target electron scattering
$e+N\rightarrow \mu +N$ at energies just above muon threshold
\cite{Diener:2004kq}.
The advantage of such a scattering experiment lies in the simplicity
of both the experiment itself and the theoretical description
which, unlike $\mu - e$ conversion,
does not depend on addressing the problems of the QED
muon-nucleus bound state. The viability of such a proposal
depends on the estimate of the $e \rightarrow \mu $ conversion cross-section
in fixed target electron scattering.
In \cite{Diener:2004kq} this has been estimated
in two scenarios involving the simplest extensions
of the Standard Model consistent with neutrino mass,
namely the case of three light Dirac neutrinos and the see-saw
mechanism \cite{Minkowski:sc}.
It was shown that even in the most optimistic scenario
with very light right-handed neutrino masses, the cross-section
is at most of order $10^{-30}$ pb, many orders of magnitude below
the capabilities of experimental detection.

The purpose of this note is to present a first calculation of the
matrix element for $e \rightarrow \mu $ conversion
within the framework of the MSSM. Technically the calculation
resembles the classical matrix element for $\mu\to e\gamma$,
but differs in that the photon in $e \rightarrow \mu $ conversion
is now off-shell, leading to a rather more complicated form
of the matrix element. One would expect the resulting
cross-section for $e+N\rightarrow \mu +N$ in the MSSM to be
many orders of magnitude larger than in non-supersymmetic
models, for the reasons discussed, and this is indeed the case.
We find that the process of $e \rightarrow \mu $ conversion
in low-energy electron-nucleus scattering provides a potentially
observable indirect signal of supersymmetry, and we therefore
advocate that such experiments be seriously considered.

\subsection{The Amplitude for the $e\to \mu$ Conversion in
  Supersymmetry (SUSY)}
The SUSY matrix element for the electron to muon conversion
at low energy is dominated by the photon exchange
with the nucleus just like the similar estimate in the
Standard Model \cite{Diener:2004kq}.
The scalar amplitude due to the Higgs exchange is
suppressed not only due to the heavy Higgs masses but also
due to the small Yukawa couplings.
The matrix element respecting Lorentz symmetry then assumes
the form
\bea
  \langle\mu|j^\lambda(0)|e\rangle
     =
  \overline{u}(p^\prime)
     &[&
           ik_\nu\sigma^{\lambda\nu}(A+B\gamma_5)
          + \gamma^\lambda           (C+D\gamma_5)
\nonumber \\
 &+&   k^\lambda (E+F\gamma_5)
     \ \ ]
          \ \ u(p)
\label{eq:me1}
\eea
where the electron, muon and photon momenta are $p$, $p^\prime$
and $k$, respectively, with $p=p^\prime + k$.
Two of the six formfactors above can be eliminated due to electric
current conservation, $k_\lambda j^\lambda = 0$. For the on-shell
electron and muon, and neglecting the electron mass, the
matrix element (\ref{eq:me1})
can then be reexpressed as
\bea
  \langle\mu|j^\lambda(0)|e\rangle
   \;\;=
   \phantom{X}\!\!\!\!\!\!
   &\phantom{+}&
  F_1\:
  \oil{u}(p^\prime)
     \left[
            \gamma^\lambda + k^\lambda\,\frac{m_\mu}{k^2}
     \right]
          u(p)
\nonumber\\
   \phantom{X} &+&
  F_2\:
  \oil{u}(p^\prime)
     \left[
            \gamma^\lambda + k^\lambda\,\frac{m_\mu}{k^2}
     \right]
          \gamma_5\,
          u(p)
\nonumber\\
   \phantom{X} &+&
  F_3\:
  \oil{u}(p^\prime)
     \left[
            \gamma^\lambda -\frac{(p+p^\prime)^\lambda}{m_\mu}
     \right]
          u(p)
\nonumber\\
   \phantom{X} &+&
  F_4\:
  \oil{u}(p^\prime)
     \left[
            \gamma^\lambda -\frac{(p+p^\prime)^\lambda}{m_\mu}
     \right]
          \gamma_5\,
          u(p),
\nonumber \\
&\phantom{X}&
\eea
where the formfactors
   $F_1 = C = -(k^2/m_\mu)E$,
   $F_2 = D =  (k^2/m_\mu)F$,
   $F_3 = m_\mu A$
and
   $F_4 = -m_\mu B$
originate from loop diagrams that violate lepton flavour and depend
on the particle masses involved in the loops.
If low-energy supersymmetry is the answer to the hierarchy problem
the formfactors emerge at the scale
${\mathcal O}(100\,$GeV$\,)$. Above this scale the effective lagrangian
is the MSSM lagrangian with small lepton-flavour violating
terms whose origin is explained in a more complete theory at a scale
$\mu_{LFV}\gg 100\,$GeV. The relevant loops contain chargino-sneutrino,
neutralino-selectron, and lepton-Higgs exchanges.

In our calculation presented below
we neglect the Yukawa couplings of the electron and muon
in order to obtain a simple and transparent result for the formfactors
$F_i$. Furthermore, to focus on the single dominant contribution we
neglect the hypercharge gauge coupling and off-diagonal chargino
and neutralino masses. We thus identify the wino-slepton loop
as the dominant contribution to the electron-to-muon conversion.
This approximation will provide an excellent estimate in the case when
the higgsino mass (the $\mu$ term) is much greater than the gaugino
mass $M_2$, say $M_2={\mathcal O}(100\,$GeV$)$ and
$\mu={\mathcal O}(1\,$TeV$)$,
as often happens in the low tan$\beta$ regime in the constrained
MSSM.
It also reduces the number of the independent formfactors
since it is the left electrons that dominate this contribution
and we can write
$\langle\mu|j^\lambda(0)|e\rangle = F_+{\mathcal M}_+ + F_-{\mathcal M}_-$
where
   $F_+=F_1=-F_2$,
   $F_-=F_3=-F_4$,
   ${\mathcal M}_+ =
                     \oil{u}(p^\prime)
     \left[
            \gamma^\lambda + k^\lambda\,\frac{m_\mu}{k^2}
     \right]
          u_L(p)$
and
   ${\mathcal M}_- =
                     \oil{u}(p^\prime)
     \left[
            \gamma^\lambda -\frac{(p+p^\prime)^\lambda}{m_\mu}
     \right]
          u_L(p)$.
Next, we present the results for the formfactors $F_+$ and $F_-$.

\begin{figure*}[t]
\scalebox{1.00}{
                \includegraphics{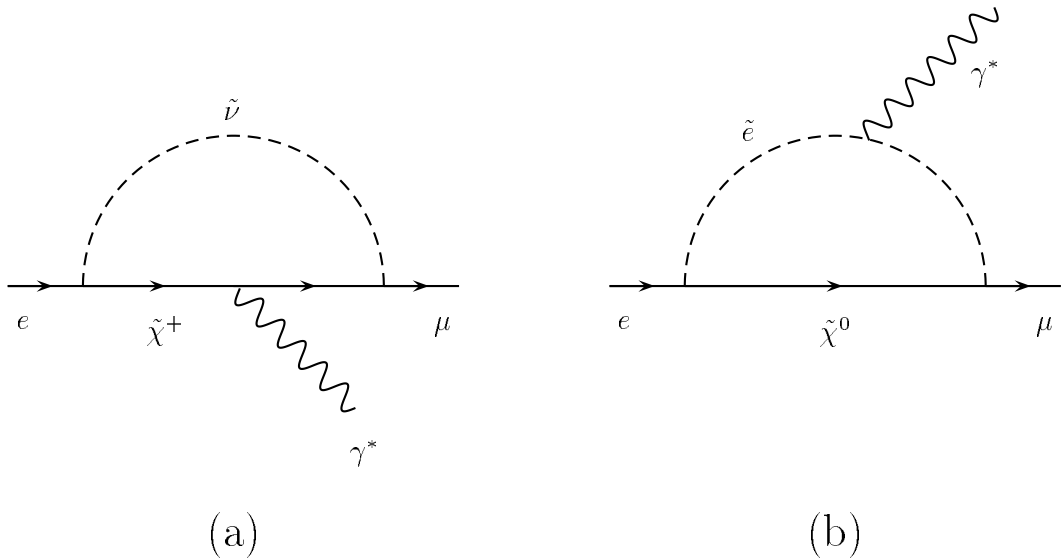}
               }
\caption{
         Feynman diagrams for the dominant MSSM contribution
         to the process $e\to \mu\gamma^*$.
}
\label{fig:feyn.diag}
\end{figure*}

The MSSM interaction of the wino-like charginos and neutralinos
with charged leptons is given by
\be
   {\mathcal L}
                       =
                         \oil{e_i}_{L}
                            \left[
                               -g_2\,
                               (\Gamma^\dagger_{\nu\,L})_{i\alpha}\:
                               \tilde{w}^-\,\tilde{\nu}_\alpha
                               +
                               \frac{g_2}{\sqrt{2}}\,
                               (\Gamma^\dagger_{e\,L})_{i\beta}\:
                               \tilde{w}^0\,\tilde{e}_\beta
                            \right]
                          + h.c.,
\label{eq:L}
\ee
where $i=1,2$ corresponds to the electron and muon, respectively, and
the $\Gamma$ matrices describe the additional slepton mixing after
the sleptons have been rotated the same way as the charged leptons.
The chargino loop, figure \ref{fig:feyn.diag}a, with the photon
attached to the fermionic line,  results in
\bea
    F_+^{(+)} &=& \frac{g^2 e}{16\pi^2}\:
             (\Gamma^\dagger_{\nu\,L})_{2\alpha}\,
             (\Gamma_{\nu\,L})_{\alpha 1}\:
             \frac{k^2}{m^2_\alpha}\:
             \left[
                     \frac{1}{3}\, f_1
                    +\frac{1}{3}\, f_3
             \right],
\nonumber\\
    F_-^{(+)} &=& \frac{g^2 e}{16\pi^2}\:
             (\Gamma^\dagger_{\nu\,L})_{2\alpha}\,
             (\Gamma_{\nu\,L})_{\alpha 1}\:
             \frac{m_\mu^2}{m^2_\alpha}\:
             \left[
                               f_1
             \right],
\eea
while the neutralino loop, figure \ref{fig:feyn.diag}b,
with the photon attached to the scalar, gives
\bea
    F_+^{(0)} &=& \frac{g^2 e}{16\pi^2}\:
             (\Gamma^\dagger_{e\,L})_{2\alpha}\,
             (\Gamma_{e\,L})_{\alpha 1}\:
             \frac{k^2}{m^2_\alpha}\:
             \left[
                    -\frac{5}{6}\, f_1
                    +\frac{1}{6}\, f_3
             \right],
\nonumber\\
    F_-^{(0)} &=& \frac{g^2 e}{16\pi^2}\:
             (\Gamma^\dagger_{e,L})_{2\alpha}\,
             (\Gamma_{e\,L})_{\alpha 1}\:
             \frac{m_\mu^2}{m^2_\alpha}\:
             \left[
                     \frac{1}{6}\, f_1
                    -\frac{1}{12}\,f_3
             \right].
\nonumber\\
&\phantom{X}&
\eea
The $f_i$ functions are
\bea
     f_1&=& \frac{1}{12(x-1)^4}\:\left(
                 x^3-6x^2+3x+2+6x\log x \right),
\nonumber\\
     f_3&=& \frac{1}{2(x-1)^3}\:\left(
                 x^2-4x+3+2\log x \right)
\eea
where $x = M_2^2/m_\alpha^2$, the ratio of the wino
mass squared over the slepton mass squared.

It is instructive to consider the limit $m_\alpha^2\ll M_2^2$.
In this case the sum over the slepton eigenstates simplifies
giving the estimate for the total SUSY contribution
\bea
    F_+ &=& -\frac{g^2 e}{16\pi^2}\:
             \frac{k^2}{M^2_2}\:
             \frac{1}{6}\:
             (m^2_L)_{21},
\nonumber\\
    F_- &=& -\frac{g^2 e}{16\pi^2}\:
             \frac{m_\mu^2}{M^2_2}\:
             \frac{7}{72}\:
             (m^2_L)_{21},
\eea
where $F_{\pm} = F_{\pm}^{(+)} + F_{\pm}^{(0)}$
and $(m^2_L)_{21}$ is the off-diagonal entry in the mass
matrix for the slepton doublet in the basis where sleptons
have been rotated by the same rotation that diagonalises the
charged leptons: $(m^2_L)_{21} = (\Gamma^\dagger_{\ell\,L})_{2\alpha}\,
                                 m_\alpha^2\,
                                 (\Gamma_{\ell\,L})_{\alpha 1}$.


\subsection{The Cross-section and
         Its Correlation with the $\mu\to e\gamma$ Constraint  }

Following \cite{Diener:2004kq} we write:
\be
    \frac{d\sigma(e+N\to \mu+N)}{d\Omega}
                 =
                     \frac{(Ze)^2}{(4\pi)^2}\,\frac{1}{k^4}
                              \:|\langle\mu|j^\lambda(0)|e\rangle |^2,
\ee
for a pointlike nuclear charge distribution.
The integration leads to the estimate
\be
    \sigma(e+N\to \mu+N) \approx 10^{-1}
                                        \:\left(
                                    \frac{M_W}{M_2}
                                          \right)^4\,
                                        \:\left(
                                    \frac{(m_L^2)_{12}}{M_2^2}
                                          \right)^2\,fb
\label{eq:sigma1}
\ee
in the approximation $m_\alpha^2\ll M_2^2$,
for $Z\approx 70$ and electron energy $E=200$MeV.

An upper bound on the cross-section above
can be obtained from the non-observation
of $\mu\to e\gamma$ decay.
The MSSM branching ratio for this decay can be expressed as
\be
    BR(\mu\to e\gamma) = \frac{48\pi^3\alpha}{G_F^2}\;|A_R|^2.
\ee
Neglecting the small yukawa couplings and keeping only
the dominant chargino-sneutrino contribution
\be
   A_R = (-i)\,\frac{g_2^2}{16\pi^2}\,\frac{1}{m_\alpha^2}\,
                (\Gamma^\dagger_{\nu\,L})_{1\alpha}\,
                                 (\Gamma_{\nu\,L})_{\alpha 2}\:
                f_1(x_\alpha),
\ee
with the same notation as in the previous section.
For $m_\alpha^2\ll M_2^2$ the branching ratio can be approximated
as
\be
    BR(\mu\to e\gamma) \approx 10^{-4}
                               \:\left(
                               \frac{M_W}{M_2}      \right)^4\;
                               \left(
                                    \frac{(m_L^2)_{12}}{M_2^2}
                                          \right)^2
\label{BR}
\ee
where we have written $\frac{\alpha}{6\pi}\approx 10^{-4}$.
By comparing to (\ref{eq:sigma1})
the current experimental
upper limit $BR(\mu\to e\gamma)\leq 10^{-11}$
\cite{lfv} then leads to the upper bound
\footnote{In making this comparison we
assume that $(m_L^2)_{12}=(m_L^2)_{21}$ which is valid in
the case that phases are neglected.}
\be
    \sigma(e+N\to \mu+N) \leq 10^{-8}fb.
\ee

\subsection{Conclusions}

We have presented a first calculation of the matrix element
for $e\to \mu$ conversion
 in low-energy electron-nucleus scattering
within the framework of supersymmetry.
 We have estimated the rate for this process in the MSSM,
 in the approximation that the wino mass parameter $M_2$
 is much smaller than the Higgsino mass parameter $\mu$,
 so that Fig.\ref{fig:feyn.diag} approximates to two wino-slepton
 loops, one with a charged wino and one with a neutral wino.
 Taking into account the existing constraint
 from non-observation of the $\mu\to e\gamma$
 decay we find the $e\to \mu$ conversion cross-section to be approximately
bounded by $\sigma < 10^{-8}\,$fb.
The cross-section may be orders of magnitude larger than
in the simplest extensions of the Standard Model, and provides
a new and potentially observable indirect signal of supersymmetry.
We strongly urge our experimental colleagues to consider performing
such an experiment.

%
%

\end{document}